\documentclass{blois}

\def\bma#1{\mbox{\boldmath{$#1$}}}

\newcommand{\arXiv}[2]{\href{http://arxiv.org/pdf/#1}{{\tt [#2/#1]}}}
\newcommand{\arXivold}[1]{\href{http://arxiv.org/pdf/hep-#1}{{\tt [#1]}}}
\def\bma#1{\mbox{\boldmath{$#1$}}}

\def\be{\begin{equation}}
\def\ee{\end{equation}}
\def\bea{\begin{eqnarray}}
\def\eea{\end{eqnarray}}

\newcommand{\Photo}{\includegraphics[height=0.1mm]{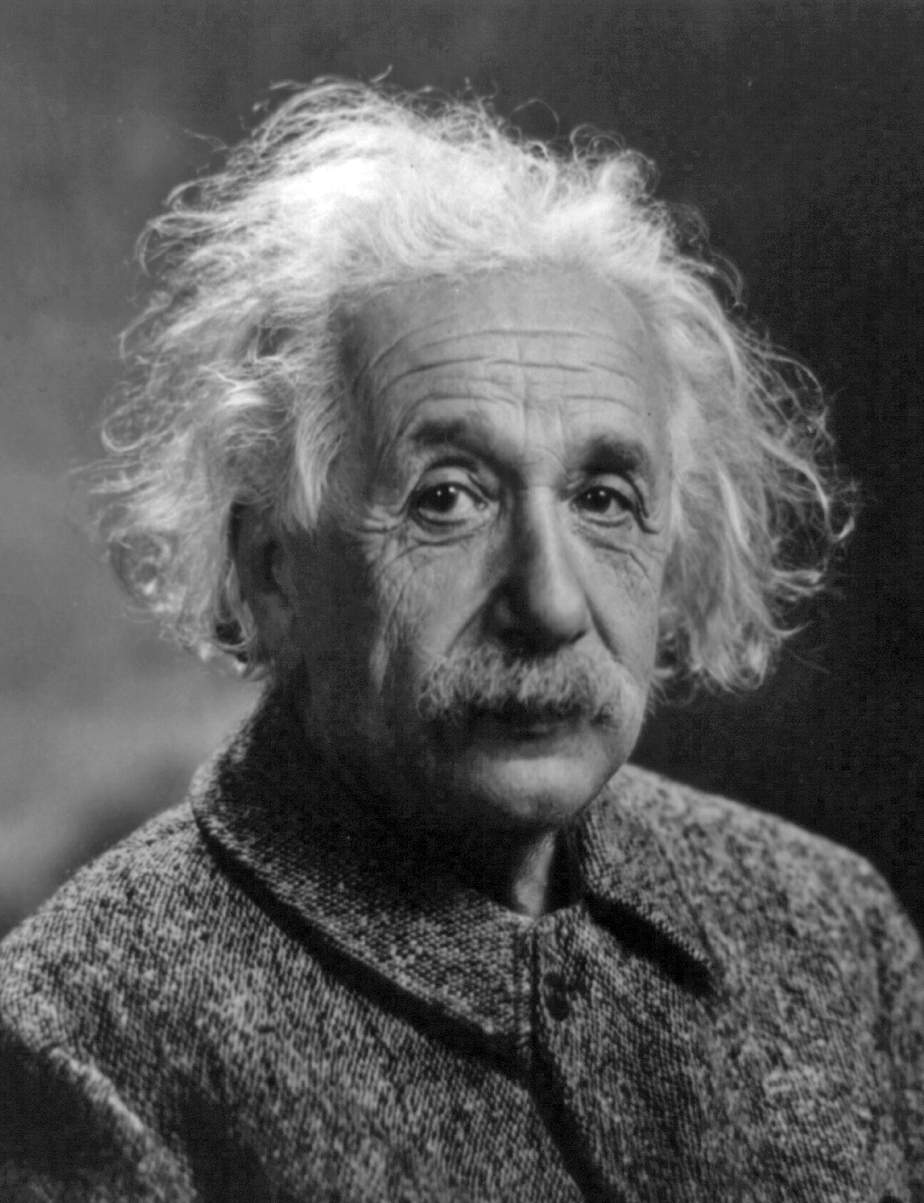}}

\begin{document}
\vspace*{4cm}
\title{A Unified View of Vacuum Decay Channels}

\author{J.R. Espinosa}

\address{Instituto de F\'{\i}sica Te\'orica UAM/CSIC, \\ 
C/ Nicol\'as Cabrera 13-15, Campus de Cantoblanco, 28049, Madrid, Spain}

\maketitle\abstracts{The tunneling potential formalism, an alternative to the standard Coleman Euclidean  approach, offers in a natural way a unified view of vacuum decays. In particular, I show in this talk how Coleman's bounce is just a member of a continuous family of solutions that can also feature Hawking-Moss instantons, pseudo-bounces and  bubble-of-nothing solutions.}

\section{The Tunneling Potential Formalism}

The tunneling potential formalism \cite{E,Eg} calculates the tunneling action for vacuum decay [of the false vacuum at $\phi_+$ of  some scalar potential $V(\phi)$]
by minimizing the simple functional
\be
\label{SVt}
S[V_t]=\frac{6\pi^2}{\kappa^2}\int_{\phi_+}^{\phi_0}\frac{(D+V_t')^2}{V_t^2D}d\phi\ .
\ee
The function $V_t(\phi)$ is the tunneling potential that connects $\phi_+$ to some field value $\phi_0$ ($>\phi_+$ by convention) beyond the potential barrier that would stabilize classically $\phi_+$. The formula above includes gravitational corrections to the decay: $\kappa=1/m_P^2$, with $m_P$ the (reduced) Planck mass, and $D=\sqrt{V_t'{}^2+6\kappa(V-V_t)V_t}$. This formalism reproduces the standard results obtained using the Euclidean methods by Coleman and de Luccia \cite{Coleman,CdL} (CdL).

The minimization procedure returns a specific solution $V_t$ that is a solution of the Euler-Lagrange equation $\delta S/\delta V_t=0$ associated to (\ref{SVt}) that reads
\be
\label{EoMVt}
(4V_t'-3V')V_t'+6(V-V_t)[V_t''+\kappa(3V-2V_t)]=0\ .
\ee
The solution extends from $\phi_+$ to some $\phi_0$ that coincides with the central value of the CdL bounce. For Minkowski or AdS false vacua, $V_t$ is monotonically decreasing (as it goes away from $\phi_+$) while for dS vacua, $V_t=V$ for some field interval from $\phi_+$ to some intermediate $\phi_i$ and, from $\phi_i$  to $\phi_0$, $V_t$ is concave down (this second field range coincides with that of the CdL bounce). For large enough $V(\phi_+)>0$, the CdL bounce does not exist and vacuum decay proceeds via the Hawking-Moss instanton \cite{HM} (with the field jumping to the top of the barrier homogeneously inside the dS horizon)  with a specific rate ($S_{HM}=24\pi^2[1/V(\phi_+)-1/V(\phi_T)]/\kappa^2$) that is reproduced by (\ref{SVt}) for $V_t=V$ from $\phi_+$ to the top of the barrier at $\phi_T$. 

\section{A Family of $\bma{V_t}$ Solutions}
The $V_t$ corresponding to the CdL solution turns out to be one (special) member of a larger family of solutions of the differential equation (\ref{EoMVt}) which also have physical meaning, as shown below. Let us consider the case of a dS false vacuum to show this  and let us look for solutions of (\ref{EoMVt}) which, like the CdL one, have some interval with $V_t=V$ (the trivial HM regime) and some interval with $V_t\leq V$ (the non-trivial regime). Calling as before $\phi_i$ the field value that separates both regimes, the family of solutions is labeled by the $\phi_i$ value (which runs from
the false vacuum $\phi_+$ to $\phi_T$, the field value at the top of the potential barrier). The boundary conditions at $\phi_i$ are as for the CdL solution:
\be
V_t(\phi_i)=V(\phi_i)\ , \quad V_t'(\phi_i)=3V'(\phi_i)/4 \ .
\ee
To find the family of solutions, ones solves (\ref{EoMVt}) starting from $\phi_i$ with these boundary conditions until the solution hits $V$ (at some end-point $\phi_e$) or diverges to $-\infty$.

\begin{figure}
\begin{minipage}{0.33\linewidth}
\centerline{\includegraphics[width=\linewidth]{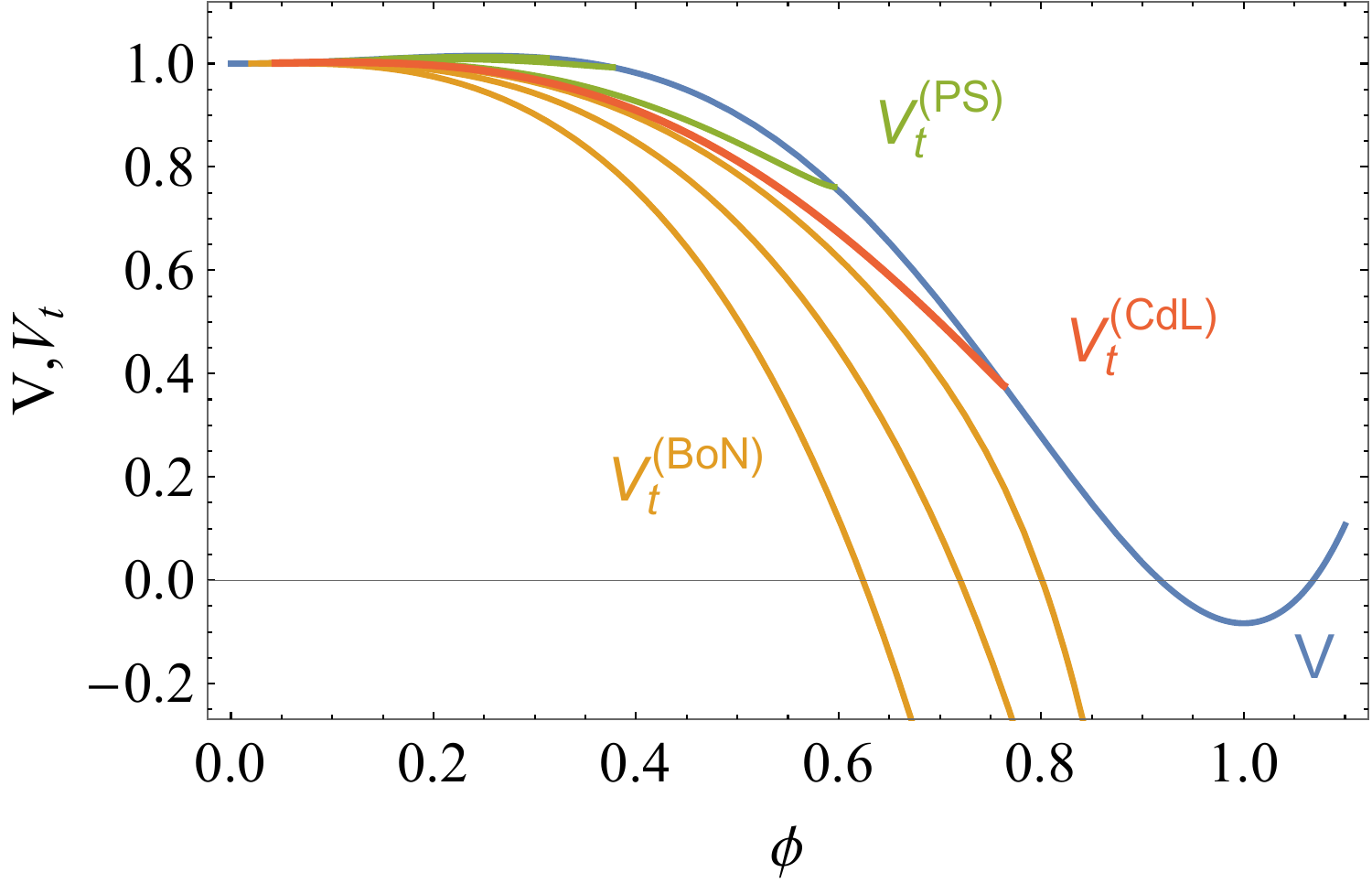}}
\end{minipage}
\hfill
\begin{minipage}{0.32\linewidth}
\centerline{\includegraphics[width=\linewidth]{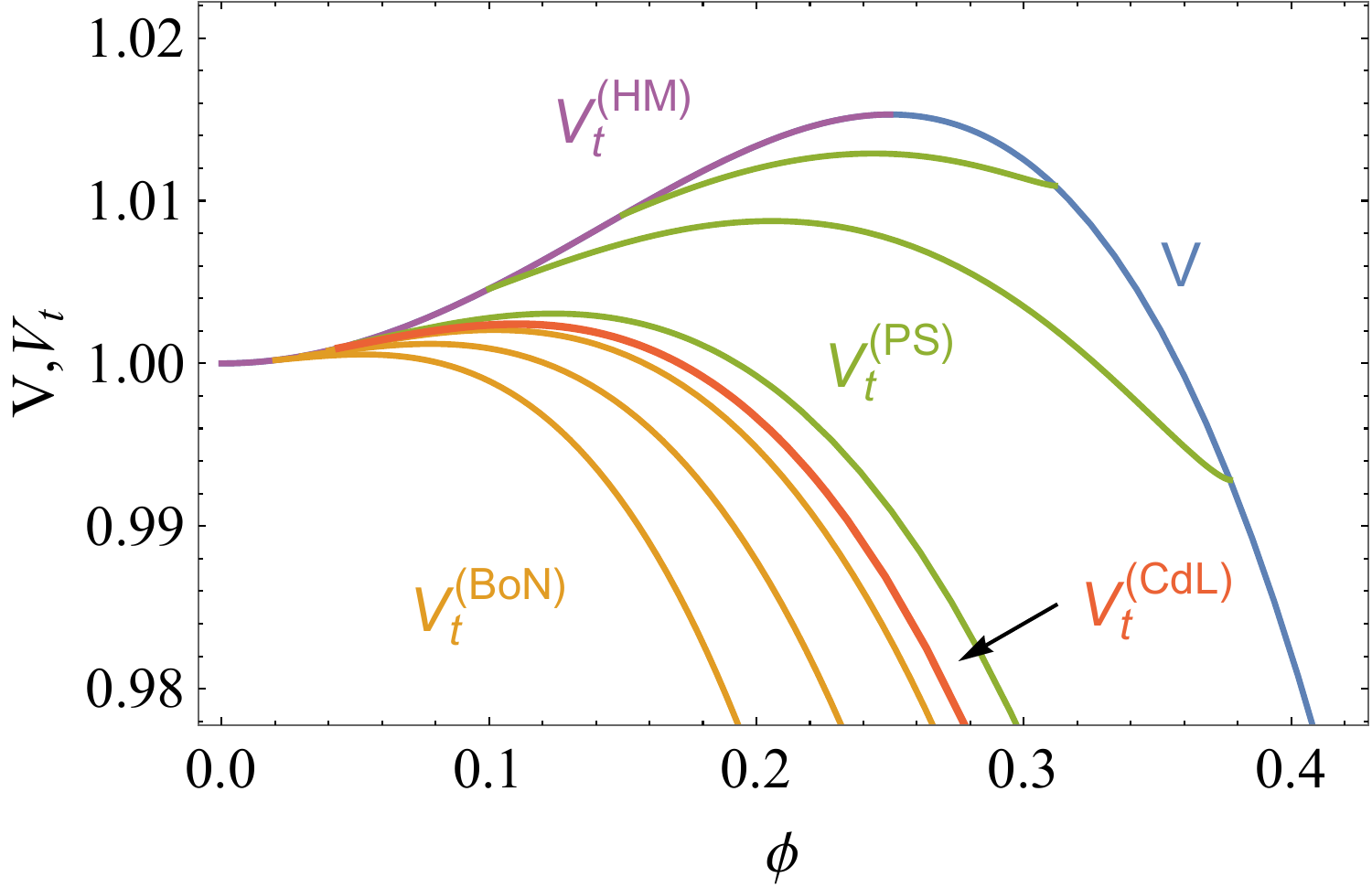}}
\end{minipage}
\hfill
\begin{minipage}{0.32\linewidth}
\centerline{\includegraphics[width=\linewidth]{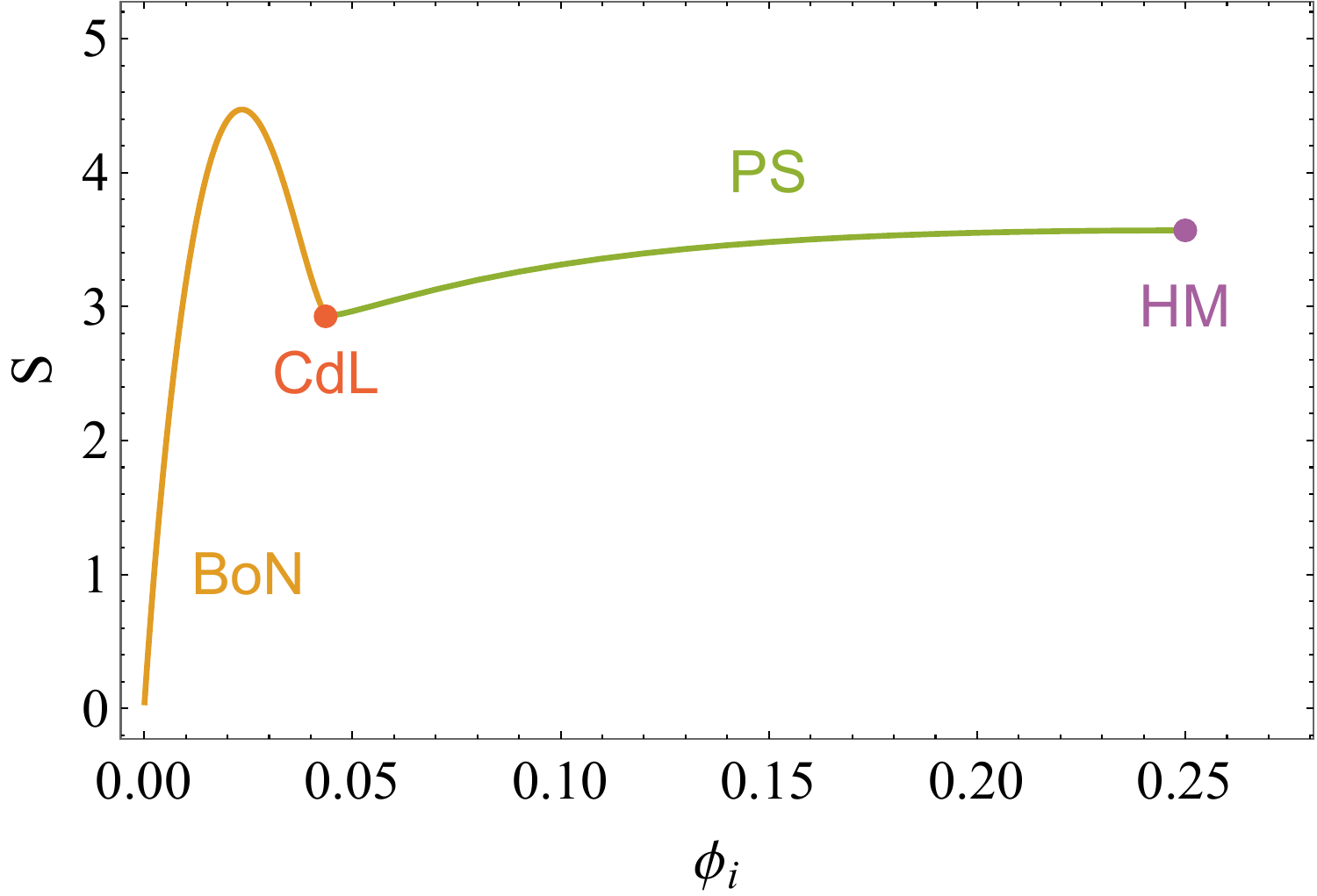}}
\end{minipage}
\caption[]{Left: Family of $V_t$ solutions for a dS false vacuum decay. Center: zoom of previous. The family includes the Hawking-Moss solution (violet), pseudo-bounces (green), the CdL solution (red) and bubbles-of-nothing (orange).  Right: Tunneling action corresponding to the different $V_t$ solutions as a function of $\phi_i$.}
\label{fig:dS}
\end{figure}

The left and central plots of figure \ref{fig:dS} show the family of solutions obtained. For $\phi_i=\phi_T$ (the top of the barrier, $\phi_T\simeq 0.25$ in the example) one gets the trivial HM solution (violet line). For smaller $\phi_i$ one gets solutions (green lines) that correspond to the so-called pseudo-bounces \cite{PS}, to be discussed in section \ref{PS}. For some special value of $\phi_i$ this family of pseudo-bounces ends, at the CdL solution (in red). Below the CdL line, the $V_t$ solutions (orange lines) diverge to $-\infty$ without hitting $V$. These correspond to bubble-of-nothing (BoN) decays \cite{WittenBoN} (provided $\phi$ is a geometrical modulus describing the size of an extra dimension), see section \ref{BoN}.

The right plot of figure \ref{fig:dS} gives the tunneling action of all these $V_t$ solutions (following the same color coding), which is finite even for the singular BoN ones. For the solutions relevant for standard (non-moduli) scalar fields (HM, pseudo-bounces and CdL) the HM and CdL ones are stationary points of the action, with CdL being the minimum, as expected.

\section{Pseudo-bounces\label{PS}}
Pseudo-bounces are vacuum decay channels that are not stationary points of the action (as seen in the example above, figure \ref{fig:dS}, right plot) but they would minimize the action $S[V_t]$
if the end point $\phi_e$ were to be held fixed. Therefore, these solutions are in principle of subleading importance for decay compared to the CdL/HM ones. However, they can become relevant \cite{PS} in at least two cases: {\it a)} if the slope of the action is small, the action for pseudo-bounces is not much larger than that of the CdL solution and one should integrate over them in order to get a good estimate of the vacuum decay rate; {\it b)} if the potential is such that the CdL solution is only reached at $\phi_e\to\infty$ (so that the CdL solution does not properly exist)
then vacuum decay necessarily proceeds via pseudo-bounce configurations. One can think of the pseudo-bounce solutions as lying along the bottom of a valley of the tunneling action in configuration space. Indeed, an slice of such valley at fixed $\phi_e$ (the central value of the field configuration) is minimized by the pseudo-bounce that ends at $\phi_e$.

\begin{figure}
\begin{center}
\begin{minipage}{0.33\linewidth}
\centerline{\includegraphics[width=\linewidth]{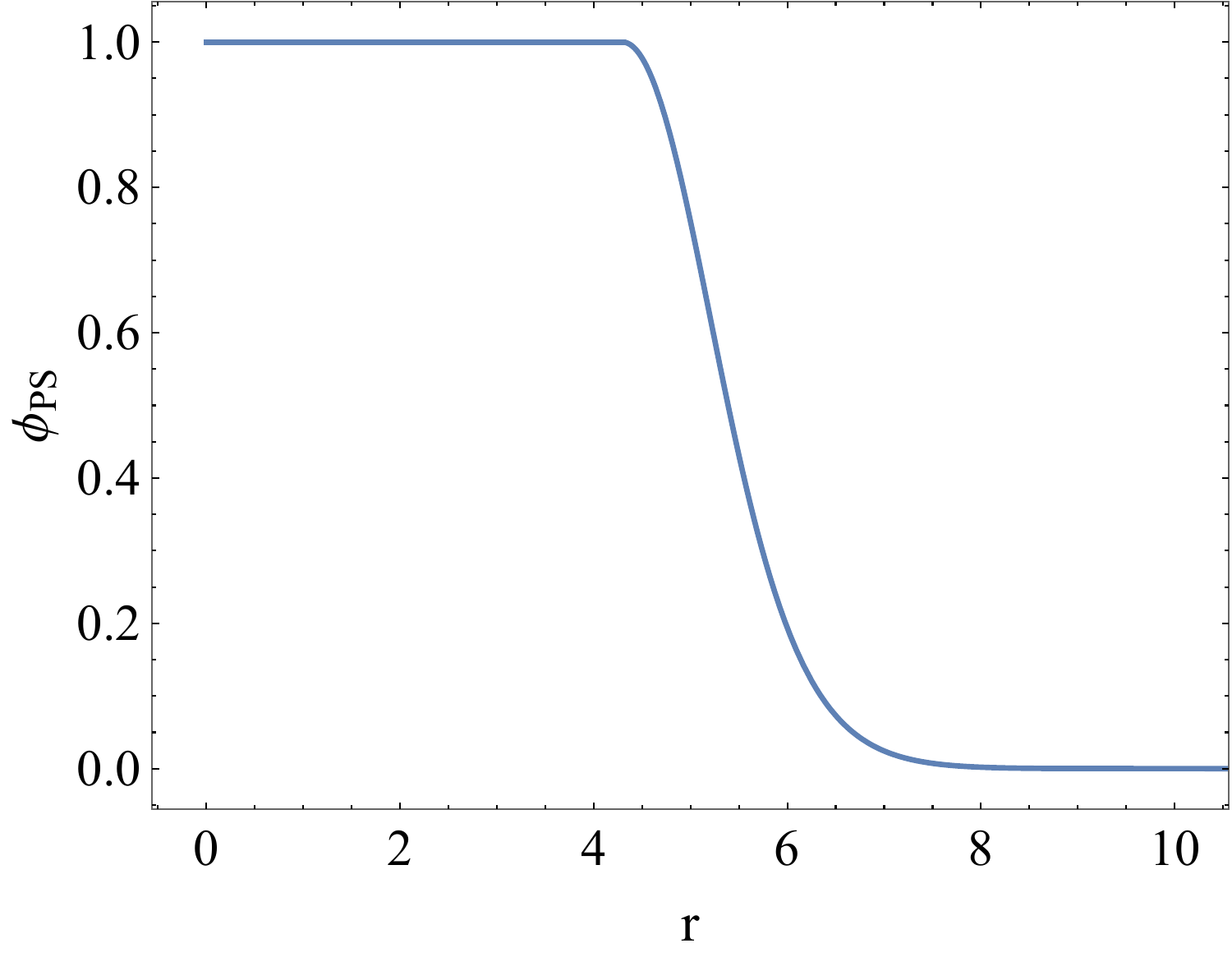}}
\end{minipage}
\begin{minipage}{0.4\linewidth}
\centerline{\includegraphics[width=\linewidth]{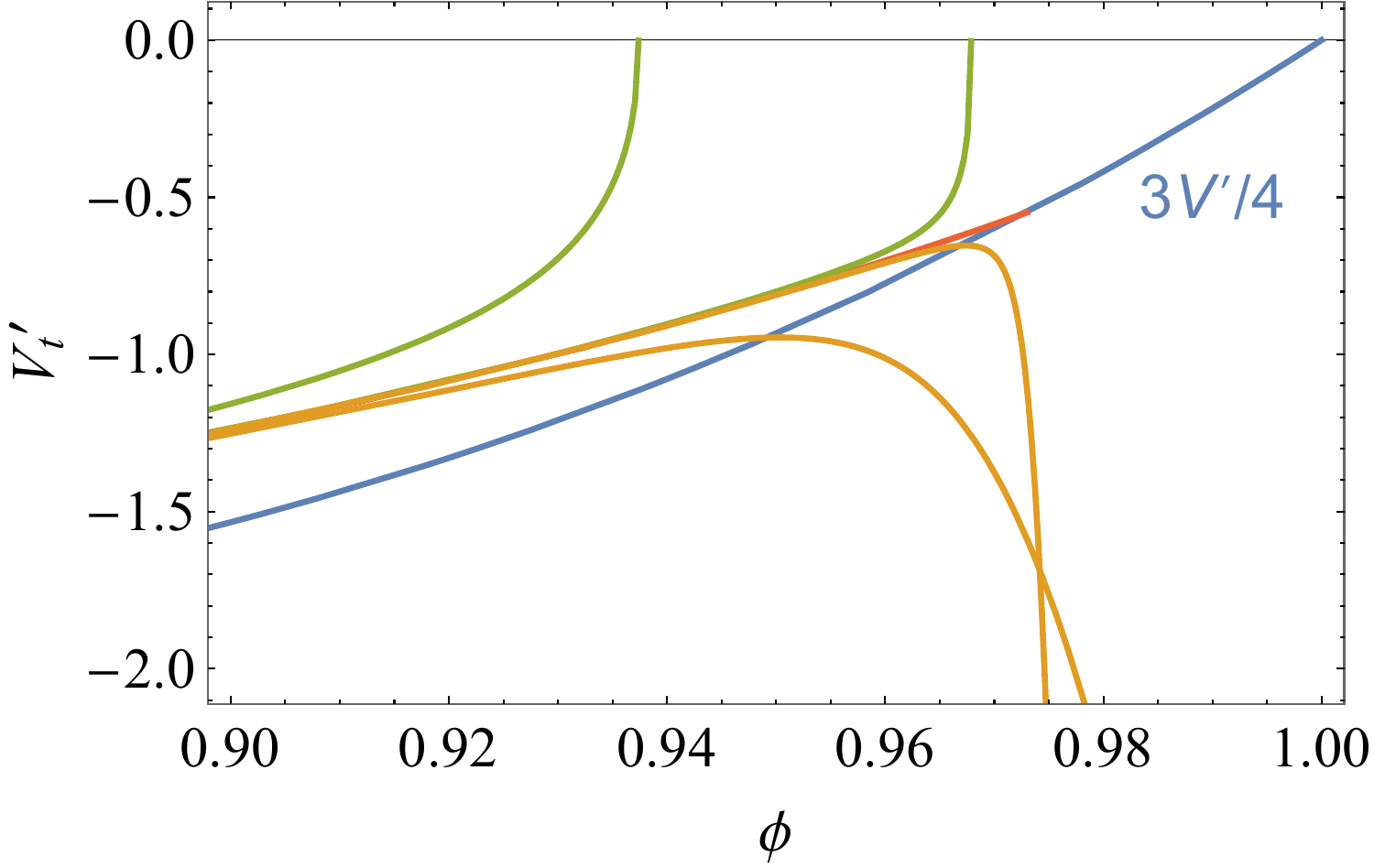}}
\end{minipage}
\end{center}
\caption[]{Left: Euclidean pseudo-bounce field profile. Right: Flow of $V_t'$ near the end point $\phi_e$ for $V_t$ solutions near the CdL one (red), which ends with $V_t'=3V'/4$. Pseudo-bounce solutions (green) flow to $V_t'=0$ while BoN solutions (orange) diverge to $V_t'=-\infty$.}
\label{fig:Vtp}
\end{figure}

The Euclidean field profile of pseudo-bounces is readily found \cite{PS}. They are configurations with a fixed field-value in their core, which extends out to some finite radius, after which the solution relaxes to the false vacuum (solving in that region the Euclidean bounce equation), see figure \ref{fig:Vtp}, left plot. In the $V_t$ formulation, the difference between the CdL solution and pseudo-bounces is that at $\phi_e$ one has $V_t'(\phi_e)=3V'(\phi_e)$ for the CdL case and $V_t'(\phi_e)=0$ for pseudo-bounces. This difference comes about naturally when solving  (\ref{EoMVt}) starting from $\phi_i$, as illustrated by figure \ref{fig:Vtp}, which shows in its right plot how $V_t'$ behave as $V_t$ approaches $V$ for the different types of solutions: for pseudo-bounces, $V_t'$ flows to 0; for CdL,
$V_t'$ flows to the finite value $3V'(\phi_e)/4$, and, for BoNs, $V_t'$ diverges to $-\infty$.

\section{Bubbles of Nothing\label{BoN}}

Bubbles of nothing, first discussed by Witten \cite{WittenBoN} for the $M^4\times S^1$ Kaluza-Klein (KK) model ($4d$ Minkowski times a circle), are a qualitatively different decay channel for theories with compact extra dimensions. A BoN describes the tunneling from a homogeneous spacetime (with constant size of the extra dimension) to an spacetime with a nucleated bubble/hole where the size of the compact dimension goes to zero at the surface of the bubble. Inside the bubble there is no spacetime (thus the name) and, after nucleation, the BoN expands and destroys the original spacetime.

In Euclidean language, Witten's BoN is described by  the instanton metric
$ds^2= (1-\mathcal{R}^2/r^2)dr^2+r^2d\Omega_3^2 + R_{KK}^2 (1-\mathcal{R}^2/r^2)d\theta^2$,
where $R_{KK}$ is the KK radius, $\mathcal{R}$ is the size of the nucleated bubble,  $r\in (\mathcal{R},\infty)$, and $\theta\in (0,2 \pi)$ parametrises the KK circle. 
Continued to Lorentzian signature,\cite{WittenBoN}  it describes the tunneling from the homogeneous $M^4\times S^1$ to a spacetime with a bubble at which the radius of the 5th dimension shrinks to zero (as $r\to \mathcal{R}$). This BoN ``hole''  then expands destroying the KK spacetime.   
The BoN decay rate per unit volume  is $\Gamma/V\sim e^{-\Delta S_E}$, with $\Delta S_E=(\pi m_P R_{KK})^2$. 

BoNs admit an effective $4d$ description in terms of singular Coleman-De Luccia  
Euclidean bounces of the modulus field $\phi$ that controls the size of the compactification \cite{DFG}. This bottom-up approach is very useful to study how the potential $V(\phi)$,  needed  to stabilize $\phi$, impacts the existence and properties of BoNs \cite{DGL}. Such $4d$ reduction opens up the possibility of using tunneling potentials for BoNs, which turn out to be unbounded in the region where the extra dimension shrinks to zero. The $V_t$ approach is very efficient to explore which types of BoN are possible and it was used in Ref.~\cite{BoN} to identify four BoN types. These types are characterized by their asymptotics at $\phi\to\infty$ (the BoN core) and correspond to different numbers of extra dimensions and the possible presence of a UV defect at the BoN core.

The $5d$ BoN can be described in $4d$  by integrating the theory over the 5th dimension $\theta$, introducing the modulus scalar $\phi$ with $e^{-2\sqrt{2/3}\,\phi}\equiv 1-R^2_{KK}/r^2$ and using a Weyl rescaling to put the BoN metric in CdL form \cite{DFG}. This maps the $5d$ instanton
into a CdL metric and a field profile, $\phi(\xi)$, with the BoN core at $\phi\to\infty$ ($\xi\to 0$) and the KK vacuum at $\phi\to 0$ ($\xi \to \infty$), where $\xi$ is a rescaled version of $r$. 
The BoN CdL solution is not of standard form as the field diverges at 
$\xi=0$ but its Euclidean action is finite and equal to Witten's (after including a boundary term of $5d$ origin). In the $V_t$ approach, Witten's BoN is given by the simple tunneling potential \cite{BoN}
\be
V_t(\phi) = -
(6/R^2_{KK})
\sinh^3(\sqrt{2/3}\,\phi)\ ,
\label{WBoN}
\ee 
with $V_t(0)=0$ and $V_t(\phi\rightarrow\infty)\sim - e^{\sqrt{6}\phi}$ ,
so that $V_t$ indeed diverges to $-\infty$ at $\phi\rightarrow \infty$ (a generic property of the $V_t$'s of BoNs,  as shown in figure \ref{fig:dS}). Remarkably, the action (\ref{SVt}) in the $V_t$ formalism gives the correct result for the BoN action without adding any boundary terms.  This agreement holds in general, not only for Witten's BoN (for the proof, see Appendix~B of Ref. \cite{BoN}). 

The $V_t$ formalism is very well suited to study in a simple and direct way the generic properties of BoNs  without having to deal with field and metric profiles. The $\phi\to\infty$ asymptotics of $V_t$ and its corresponding $D=\sqrt{V_t'{}^2+6\kappa(V-V_t)V_t}$ function  depend on some parameter that cannot be calculated in the $4d$ approach (thus one gets a continuous family of BoN solutions).  Once a particular $d=4+n$ theory is considered, that parameter is fixed in terms of geometric properties of the compactified space [see (\ref{WBoN}) as one example] and one particular BoN solution is selected (thus neighboring BoN points in the action curve in figure \ref{fig:dS} correspond to different theories and it is not meaningful to consider the curve extremals). For further details, see Ref. \cite{BoN}.

\section{Conclusions}

The use of the tunneling potential approach to calculate vacuum decay solutions and their tunneling actions puts forth in a natural way a unified view of different vacuum decay channels. Focusing on dS false vacua I have shown how the Hawking-Moss instanton and the Coleman-de Luccia bounce are connected by a family of pseudo-bounce solutions and also how bubble-of-nothing solutions appear naturally beyond the CdL one. The solutions in this heterogeneous family have finite action, have a precise physical meaning and all can be relevant for decay.

The same story unfolds for Minkowski or AdS false vacua \cite{BoN}. In those cases the Hawking-Moss instanton does not exist and the pseudo-bounce family of solutions ends up at a particular solution with $D=0$ and infinite action. (For some potentials, this $D=0$ solution interpolates directly between false and true vacua and corresponds to a flat domain wall). Bubble of nothing solutions appear in a similar way below the CdL solution.

It would be interesting to explore if other decay solutions, not yet thought of, appear thanks to the application of the $V_t$ formalism.

\section*{Acknowledgments}

This work has been funded by the following grants: IFT Centro de Excelencia Severo Ochoa CEX2020-001007-S and by PID2022-142545NB-C22 funded by MCIN/AEI/10.13039/ 501100011033 and by ``ERDF A way of making Europe''.

\end{document}